\title{TT-RecS: The Taxonomic Trace Recommender System}

\documentclass[conference]{IEEEtran}

\usepackage[T1]{fontenc}
\usepackage[utf8]{inputenc}
\usepackage{booktabs}
\usepackage{tabularx}
\usepackage{graphicx}
\usepackage{subcaption}
\usepackage{todonotes}
\usepackage[bookmarks=false]{hyperref}
\usepackage{amsmath}
\usepackage{caption}
\usepackage[flushleft]{threeparttable}

\begin{document}
\author{\IEEEauthorblockN{Michael Unterkalmsteiner}
  \IEEEauthorblockA{Department of Software Engineering\\
Blekinge Institute of Technology\\
Karlskrona, Sweden\\
michael.unterkalmsteiner@bth.se}}

\maketitle

\begin{abstract}
Traditional trace links are established directly between source and target
artefacts. This requires that the target artefact exists when the trace is
established. We introduce the concept of indirect trace links between a source
artefact and a knowledge organization structure, e.g. a taxonomy. This allows
the creation of links (we call them taxonomic traces) before target artefacts
are created. To gauge the viability of this concept and approach, we developed a
prototype, TT-RecS, that allows to create such trace links either manually or
with the help of a recommender system.
\end{abstract}

\begin{IEEEkeywords}
  Traceability, Requirements, Domain-specific Taxonomy, Recommender System
\end{IEEEkeywords}

\section{Introduction}
Traceability from requirements specifications to downstream artefacts has shown
to lead to more efficient and correct software
maintenance~\cite{mader_developers_2015}, is a pre-requisite for
requirements-based testing~\cite{bouillon_survey_2013}, and in certain
application domains a necessity to demonstrate compliance to
regulations~\cite{regan_traceability-why_2012}.

However, manual creation and maintenance of trace links is in practice often not
feasible due to their number and the complexity of information that needs to be
maintained over time. Many requirements management tools support manual trace
link creation or even automated trace link recovery. However, none of them
supports, to the best of our knowledge, taxonomic traces, explained next.

\section{Taxonomic Traces}
\begin{figure*}
\begin{subfigure}{.5\textwidth}
  \centering
  \includegraphics[width=.8\linewidth]{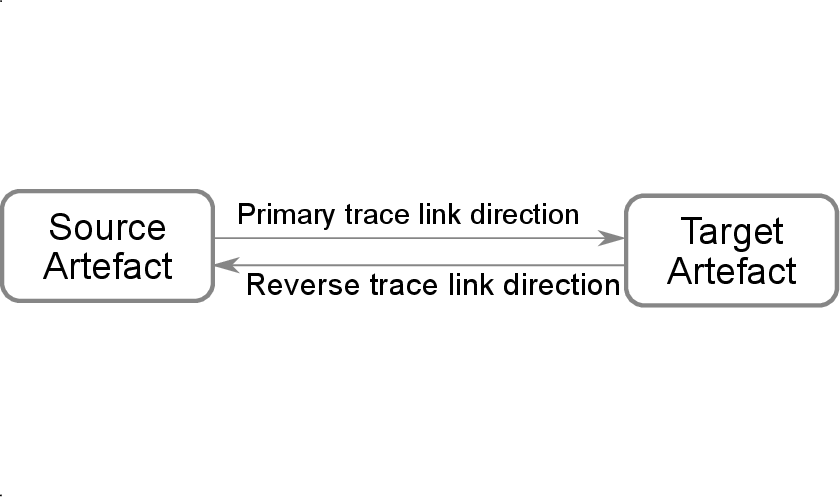}
  \caption{Traditional trace link (adapted from~\cite{gotel_traceability_2012}}
  \label{fig:traditional}
\end{subfigure}%
\begin{subfigure}{.5\textwidth}
  \centering
  \includegraphics[width=.8\linewidth]{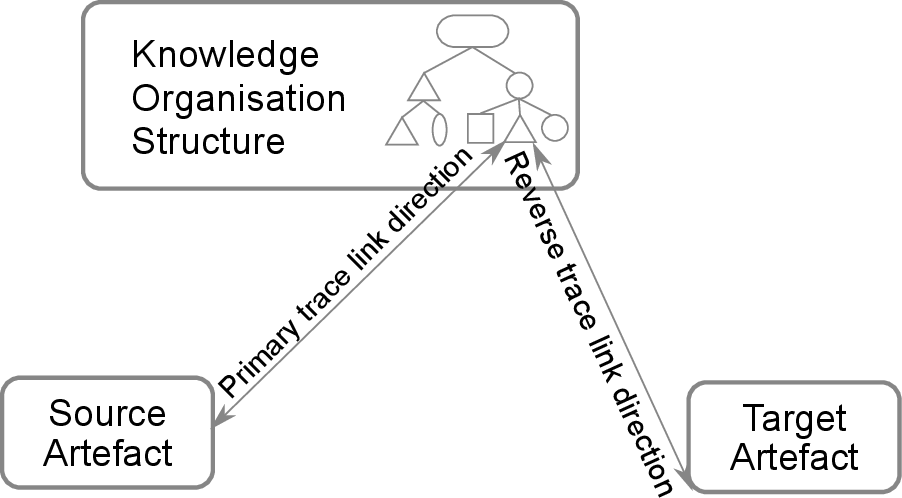}
  \caption{Taxonomic trace link}
  \label{fig:taxonomic}
\end{subfigure}
\caption{Traditional vs. taxonomic trace links}
\label{fig:fig}
\end{figure*}

In systems and software engineering, a ``trace'' is often defined as a triplet
consisting of a source artefact, a target artefact and a link associating the
two artefacts~\cite{gotel_traceability_2012} (Figure~\ref{fig:traditional}). We
propose\footnote{We do not claim novelty of this idea since, as far as we know,
  Noll and Ribeiro~\cite{noll_enhancing_2007} were first to outline in their
  position paper from 2007 on how to enhance traceability using ontologies in
  the Unified Process.}
to introduce indirect trace links to a taxonomy or similar knowledge
organization systems such as controlled vocabularies, taxonomies or ontologies (Figure~\ref{fig:taxonomic}).

Before we delve into the technical solution we prototyped with the Taxonomic
Trace Recommender System (TT-RecS) described in Section~\ref{sec:ttrecs}, we
illustrate the limitations of the traditional artefact-to-artefact trace
link approach. Conceptually, trace links connect artefacts along the following
dimensions: abstraction, structure and time.
\paragraph{Abstraction}
Traced artefacts usually exist on different abstraction levels of domain
concepts. For example, a requirement may describe the capabilities of a bank
account, while a class \verb+BankAccount+ in an object oriented design provides
an implementation. Someone (or something, e.g. an algorithm) that aims to
establish a direct trace link needs to be able to handle the difference in
abstraction.
\paragraph{Structure}
With the structure dimension we refer to how information about domain concepts
in different artefacts is stored. For example, requirements are typically stored
in tools that support natural language text (word processors or dedicated
requirements management tools). Other artefacts (design models, source code,
test cases) are stored and managed with different tool sets, specialized for the particular task.
\paragraph{Time}
The time dimension refers to the fact that artefacts typically are not created
at the same time, making it impossible to create trace links when the source
artefact is developed. For example, when a requirement is elicited and
specified, typically the implementation and test cases do not exist
yet\footnote{Test driven- and behaviour-driven development are examples of means
  to reduce this time gap.}.

We briefly outline motivations why taxonomic trace links
(Figure~\ref{fig:taxonomic}) address the above issues:
\begin{enumerate}
  \item The abstraction level gap can be reduced as engineers need only their
    domain expertise and their skills specific to their profession. This might be less
    important for a scenario with full-stack engineers of DevOps. However,
    system and civil engineers are specialized and technical domain experts,
    without necessarily being proficient in design and implementation tasks.
    \emph{Hence, experts in the problem and the solution domain respectively can
    create trace links to a common knowledge organization structure, without the
    need to tap into unfamiliar knowledge areas.}
  \item The structure gap can be bridged by a common taxonomy that connects
    information silos by removing the need of direct interoperability.
    Profession specific information management systems (for requirements, design
    documents, source code) can be adapted to support traces to knowledge
    organization structures (such as a taxonomy) instead of directly supporting
    trace links between myriads of different systems. This is especially
    important for scenarios where engineering work is outsourced, and clients
    rely on trace links to perform delivery verification. \emph{Hence, knowledge
    organization structures represent a conceptual Application Programming
    Interface (API) that information management systems can implement
    independently, enabling interoperability.} 
  \item The temporal gap is removed as engineers can create trace links at the
    same time they create the artefact, removing the necessity to recover trace
    links when downstream artefacts become available. Furthermore, the taxonomic
    trace links can be used to analyse the artefact, for example in case of
    requirements for their completeness or consistency. \emph{Hence, by
      benefiting the creators of taxonomic trace links through their
      immediate usefulness, they provide intrinsic motivation and are therefore
      more likely to be created at all.}
\end{enumerate}

While these benefits of indirect, taxonomic tracing are very promising, they have, to the best of our knowledge, not yet been evaluated in practice. We
have therefore implemented a prototype system, evaluated in a pilot experiment~\cite{unterkalmsteiner_early_2020}, and present here its basic design and functionality.

\section{Taxonomic Trace Recommender System}\label{sec:ttrecs}

\begin{figure}[b]
  \centering
  \includegraphics[width=\linewidth]{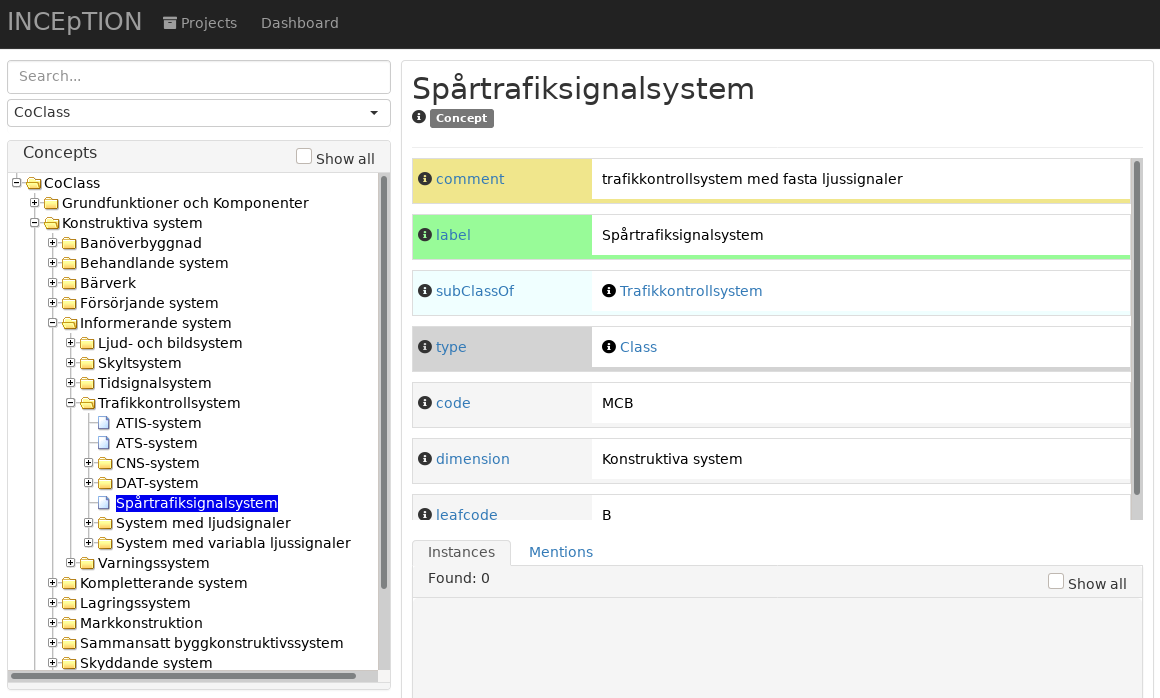}
  \captionof{figure}{Example objects from CoClass}
  \label{fig:kb}
\end{figure}

\begin{figure}[b]
  \centering
  \includegraphics[width=\linewidth]{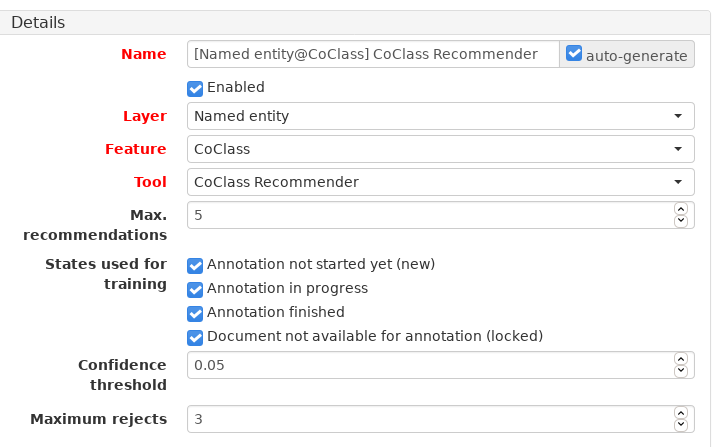}
  \captionof{figure}{Recommender settings}
  \label{fig:rs}
\end{figure}

\begin{figure*}
  \centering
  \includegraphics[width=\textwidth]{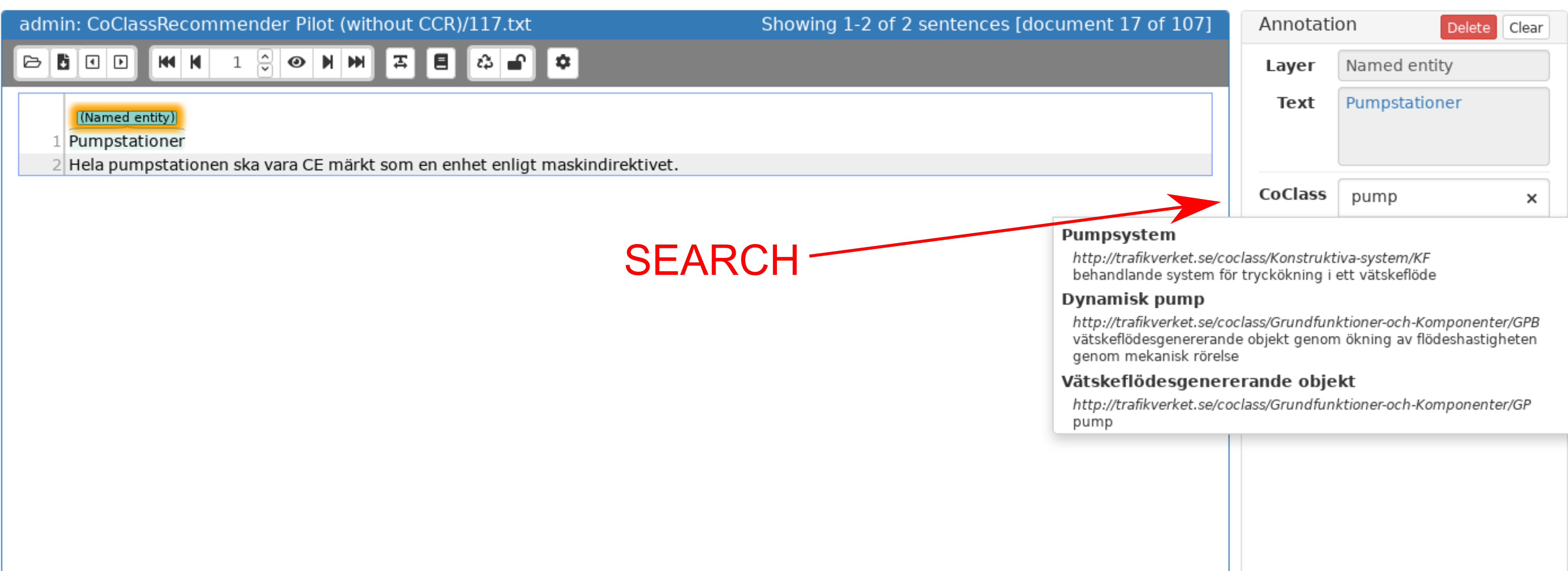}
  \captionof{figure}{Manually identified taxonomic traces through search}
  \label{fig:search}
\end{figure*}

\begin{figure*}
  \centering
  \includegraphics[width=\textwidth]{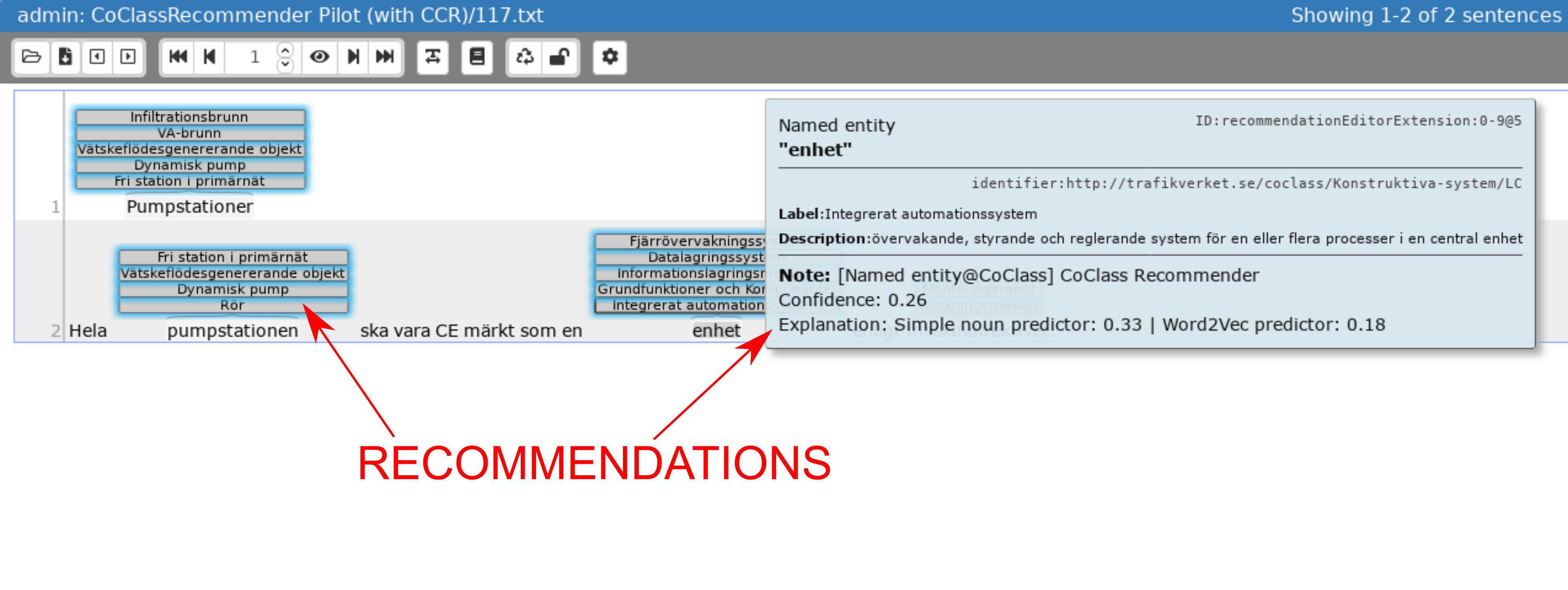}
  \captionof{figure}{Taxonomic traces identified by the recommender}
  \label{fig:recommended}
\end{figure*}

TT-RecS is built on top of INCEpTION~\cite{klie_inception_2018}, a general
purpose text annotation platform that can be extended with custom recommenders.
INCEpTION allows importing knowledge organization structures from RDF files or
connect to a remote knowledge base using SPARQL. Since we developed TT-RecS in
the context of a collaboration with Trafikverket, the Swedish Transport
Authority, we used the domain specific taxonomy
\emph{CoClass}\footnote{\url{https://coclass.byggtjanst.se/about\#about-coclass}}
which organizes concepts from the construction domain in a hierarchical
taxonomy. Figure~\ref{fig:kb} shows an example of the imported data from
\emph{CoClass} in INCEpTION's knowledge base.

The recommender configuration (Figure~\ref{fig:rs}) provides two relevant options:
\begin{enumerate}
  \item The threshold value for the recommendation confidence (a value computed
    by the different predictors explained
    in~\cite{unterkalmsteiner_early_2020}). Recommendations with a lower value
    than specified here are not shown.
  \item The maximum number of rejects after which a suggestion is not shown any more. 
\end{enumerate}

INCEpTION allows the user to import documents in different formats (plain text,
JSON, HTML, PDF and different NLP tool exchange formats).
These documents can in turn be opened in INCEpTION's annotation interface.

Figure~\ref{fig:search} shows an example where a document with a requirement is
opened without TT-RecS. The user must use his/her domain knowledge, i.e. use the
correct search terms, to find the objects from CoClass that can be associated to
the requirement in question.

Figure~\ref{fig:recommended}, on the other hand, shows the same requirement with
CoClass objects suggested by TT-RecS. The user can now either accept or reject
the suggestions, based on the information shown about the suggested CoClass
object.

We have evaluated TT-RecS in a pilot experiment with seven domain experts with
varying experience in CoClass and reading/writing
requirements~\cite{unterkalmsteiner_early_2020}. The goal of the pilot was to
validate the experiment instrument and get some indication whether domain
experts are able (with or without recommender) to create taxonomic trace links.
The main take-away of that study is that the trace task was challenging,
independently of whether the recommender was used or not. While this result was
not encouraging, we are currently improving TT-RecS and incorporate the lessons
learned from the qualitative analysis from the experiments results. We
conjecture that the main gain can be achieved by tuning the recommendations such
that they take the context of nouns into account. Since the
tracing task is challenging even for domain experts, we conjecture that
recommendation systems are essential to make taxonomic traces practicable.

\section{Demonstration Plan}
Since the concept of taxonomic traces is novel, and as far as we know not yet
implemented in a tool, we preface our demonstration with an explanation of the
concept and pointing out the differences and advantages w.r.t. traditional trace
links. Then we illustrate how to set up the demonstrator\footnote{Available
  here: \url{http://doi.org/10.5281/zenodo.3827169}} and point to the source
code of TT-RecS that is under active
development\footnote{\url{https://github.com/munterkalmsteiner/inception}}.

The demonstration proper illustrates how to configure INCEpTION with the
recommender and import documents. We then compare the manual creation of trace
links using INCEpTION's annotation interface with the recommender suggested
trace links.

\section{Conclusions}
We believe that our research will serve as a base for future studies on
taxonomic traces, in particular as we have developed a working prototype that
has been used in an experiment with industry participants. While the precision
of the recommender needs to be improved, we think it is important to evaluate
novel ideas in realistic settings, early in the ideation process, in order to
gauge the practical viability of the idea.  

\bibliographystyle{IEEEtran}
\bibliography{main}

\end{document}